\documentclass[nofootinbib]{revtex4}
\usepackage[dvips]{graphicx}
\usepackage{epsf}
\usepackage{amsmath}
\usepackage{amssymb}

\usepackage{graphicx}% Include figure files
\usepackage{dcolumn}% Align table columns on decimal point
\usepackage{bm}% bold math
\def\edth{\;\raise1.0pt\hbox{$'$}\hskip-6pt\partial\;}
\def\baredth{\;\overline{\raise1.0pt\hbox{$'$}\hskip-6pt
\partial}\;}
\def\bi#1{\hbox{\boldmath{$#1$}}}
\def\gsim{\raise2.90pt\hbox{$\scriptstyle
>$} \hspace{-6.4pt}
\lower.5pt\hbox{$\scriptscriptstyle
\sim$}\; }
\def\lsim{\raise2.90pt\hbox{$\scriptstyle
<$} \hspace{-6pt}\lower.5pt\hbox{$\scriptscriptstyle\sim$}\; }

\def\be{\begin{equation}}
\def\ee{\end{equation}}
\def\bea{\begin{eqnarray}}
\def\eea{\end{eqnarray}}

\begin{document}

\title{$CPT$ Violating Electrodynamics and Chern-Simons Modified Gravity}

\author{Mingzhe Li$^{1,2,4}$, Yi-Fu Cai$^{3}$, Xiulian Wang$^{1,2}$ and Xinmin Zhang$^{3,4}$}

\affiliation{1) Department of Physics, Nanjing University, Nanjing
210093, P.R. China}
\affiliation{2) Joint Center for Particle, Nuclear Physics and
Cosmology, Nanjing University - Purple Mountain Observatory,
Nanjing 210093, P.R. China}
\affiliation{3) Institute of High Energy Physics, Chinese Academy
of Sciences, P.O. Box 918-4, Beijing 100049, P.R. China}
\affiliation{4) Theoretical Physics Center for Science Facilities
(TPCSF), Chinese Academy of Sciences, P.R. China}

\begin{abstract}
The electrodynamics with a Chern-Simons term
$p_{\mu}A_{\nu}\widetilde{F}^{\mu\nu}$ violates Lorentz and $CPT$
symmetries with a non-vanishing $p_{\mu}$. For a fixed vector
$p_{\mu}$, in this paper we point out that the energy-momentum
tensor of this theory coupled to the gravity minimally is
symmetric but not divergence free, which consequently makes the
gravitational field equation inconsistent. To preserve the
consistency, we introduce a Chern-Simons term in the gravity
sector with the coefficient determined by the Lorentz and $CPT$
violating term in the electromagnetic field. Further we study the
phenomenologies of the model on the cosmic microwave background
radiation and the relic gravitational waves.
\end{abstract}

\maketitle
%\hskip 1.6cm PACS number(s): 98.80.Es, 98.80.Cq, 98.70.Vc \vskip 0.4cm

\section{Introduction}

Both the Standard Model of particle physics and Einstein's general
relativity are locally Lorentz and $CPT$ invariant. Probing their
violations is an important way to search for new physics and in
recent years has attracted a lot of interests. Usually, studies are
focused on the phenomenologies in matter and gravity sectors
separately. However there is an issue related to the consistency of
the theory which needs to be examined. In a class of models of
Lorentz symmetry breaking, fixed preferred frames are introduced
\cite{Carroll:1989vb, Colladay:1998fq}. The violation effects are
formulated by introducing operators like $p_{\mu} K^{\mu}$ in the
Lagrangian of the matter with $p_{\mu}$ being a fixed vector
and $K^{\mu}$ the matter current. The existence of the fixed
preferred frames violates general coordinate covariance also. If the
Lorentz invariance in the matter sector is broken in this way, it is
impossible to get an energy-momentum tensor of the matter which is
both symmetric and covariantly conserved. When gravity is included,
this makes the Einstein's equation $G^{\mu\nu}\equiv R^{\mu\nu}-1/2
Rg^{\mu\nu} =-8\pi G T^{\mu\nu}$ inconsistent. Because $G^{\mu\nu}$
is symmetric between the indices $\mu$ and $\nu$, and its divergence
is identically zero as the result of the contracted Bianchi
identity. In this paper we provide a solution to this problem by
modifying the gravity simultaneously in a non-covariant way, so that
the new terms in the gravity sector balance the Lorentz violating
effects in the matter sector.

As a concrete example, in this paper we consider a matter sector where the
electrodynamics is modified by a Chern-Simons term
$p_{\mu}A_{\nu}\widetilde{F}^{\mu\nu}$ .
 This term breaks Lorentz and $CPT$
symmetries if $p_{\mu}$ is treated as an external field and the
related phenomenology has been studied extensively in the
literature, such as testing $CPT$ of photons in astrophysics
\cite{Kostelecky:2008be} and cosmology
\cite{Lue:1998mq,Feng:2004mq,Feng:2006dp,Li:2006ss,Li:2008tma,WMAP5,xia2,Wu:2008qb,Geng:2007va}.
A salient feature of this modified electrodynamics is the rotations
of the polarizations of propagating photons. The rotation angle
$\Delta \chi$ depends on the external field. Due to this feature, a
part of $E$ type polarization will be rotated to $B$ type
polarization for photons. This will generate $TB$ and $EB$
correlations in the power spectra of the cosmic microwave background
radiation (CMB). Hence we can use the CMB experiments to test the
Lorentz and $CPT$ violation in this model. With homogeneous and
isotropic rotation angle, in Ref. \cite{Feng:2004mq}, two of us with
Feng and Li did the simulations on the measurement of $\Delta\chi$
with the CMBPol and PLANCK experiments. We pointed out that in such
experiments the $EB$ spectrum will be the most sensitive probe of
such Lorentz and $CPT$ violation. In \cite{Feng:2006dp}, two of us
with Feng, Xia, and Chen first found that a nonzero rotation angle
$\Delta\chi=-6.0\pm4.0$ deg is mildly favored by the CMB
polarization data from the three-year Wilkinson Microwave Anisotropy
Probe (WMAP3) observations
\cite{Spergel:2006hy,Page:2006hz,Hinshaw:2006,Jarosik:2006,WMAP3IE}
and the January 2003 Antarctic flight of BOOMERanG
\cite{B03,B03EE,TCGC}. With the newly released five year data
\cite{WMAP5}, the WMAP group gives $\Delta\chi=-1.7\pm2.1$ deg,
which when combined with the BoomeRang data is \cite{xia2}
$\Delta\chi=-2.6\pm1.9$ deg. The result by the QUaD collaboration
 is $\Delta\chi=0.55\pm 0.82\pm 0.5$ deg \cite{Wu:2008qb} and
most recently improved to $\Delta \chi = 0.64 \pm 0.5 \pm 0.5$ deg \cite{Brown:2009uy}.

As mentioned above, for a fixed $p_{\mu}$, the electromagnetic
Chern-Simons term is not invariant under the coordinate
transformation. The energy-momentum tensor is the same as that of
the Maxwell theory, but it is not covariantly conserved because
the equation of motion is modified. This will make the Einstein's
equation inconsistent. To preserve the  consistency, we modify the
Einstein's gravity simultaneously by introducing a gravitational
Chern-Simons term. As a result, the left hand side of the
gravitational equation is modified by a four dimensional Cotten
tensor which also has non-zero divergence and will match the
divergence of the electromagnetic field on the right hand side.
 We further in this paper
study the phenomenologies of this model on the CMB
 and relic gravitational waves.

Our paper is organized as follows: in section II, we firstly
review briefly the Maxwell-Chern-Simons theory and point out the
inconsistence when including the gravity. We then introduce a
gravitational Chern-Simons term and demonstrate how the theory
becomes consistent; in section III, we study the effects of our
model on CMB polarizations; in section IV, we analyze the
late-time evolution of relic gravitational waves and our result
show the potential signals at high-frequency regime; section V is
the summary.

\section{Chern-Simons modified electrodynamics and gravity}

The Maxwell-Chern-Simons theory is the Maxwell electrodynamics modified by a Chern-Simons term:
\bea\label{chernsimons}
\mathcal{L}_F&=&-\frac{1}{4}F_{\mu\nu}F^{\mu\nu}-\frac{1}{2}p_{\mu}A_{\nu}\widetilde{F}^{\mu\nu}\nonumber\\
&=&
-\frac{1}{4}F_{\mu\nu}F^{\mu\nu}
+\frac{\theta_1}{4}F_{\mu\nu}\widetilde{F}^{\mu\nu}~,
\eea
where $F_{\mu\nu}$ is the electromagnetic field tensor and $\widetilde{F}^{\mu\nu}=1/2\epsilon^{\mu\nu\rho\sigma}F_{\rho\sigma}$ is
its dual, $p_{\mu}=\nabla_{\mu}\theta_1$ characterizes the preferred frame and has dimension $[E]$. We use the signature $(+,~-,~-,~-)$ for the metric.
In the second line of Eq. (\ref{chernsimons}),
an integration by part is used.
To characterize a fixed preferred frame, $p_{\mu}$ is assumed to be a constant vector in spacetime, so $\theta_1=p_{\mu}x^{\mu}+C$. Here $C$
is a constant, but it only contributes a surface term to the Lagrangian and can be set to zero.
We will ignore the sources for the electromagnetic field in this paper. When considering the minimal coupling to gravity,
from the Lagrangian (\ref{chernsimons}), we get the energy-momentum
tensor
\be
T^{\mu\nu}_F=-\frac{2}{\sqrt{g}}\frac{\delta S_F}{\delta g_{\mu\nu}}~,
\ee
where $g=-det g_{\mu\nu}$. It is this energy-momentum tensor provides the source to the gravity and appears in the right-hand side of the
gravitational field equation.
Because the Chern-Simons term is a topological term, it does not depend on the metric and has no contribution to the energy-momentum tensor.
So $T^{\mu\nu}_F$ is the same as that
of the Maxwell theory,
\be\label{stress}
T^{\mu\nu}_F=\frac{1}{4}F_{\alpha\beta}F^{\alpha\beta}g^{\mu\nu}-F^{\mu\alpha}F^{\nu}_{~\alpha}~.
\ee
However, the equation of motion is modified as
\be
\nabla_{\mu}F^{\mu\nu}=p_{\mu}\widetilde{F}^{\mu\nu}~.
\ee
After making use of the equation above, we find the energy-momentum tensor is not covariantly conserved,
\be\label{divergence}
\nabla_{\mu}T^{\mu\nu}_F=-\frac{1}{4}p^{\nu}F_{\mu\alpha}\widetilde{F}^{\mu\alpha}~.
\ee
Substitute the tensor in Eq. (\ref{stress}) into the Einstein equation, it becomes
\be
G^{\mu\nu}=-8\pi G(T^{\mu\nu}_F+T^{\mu\nu}_m)~,
\ee
where $T^{\mu\nu}_m$ is the energy-momentum tensor of other matter. In this paper we assume there are no other Lorentz violations except the
Chern-Simons term in the photon sector, so $T^{\mu\nu}_m$ is symmetric and divergenceless. From Eq. (\ref{divergence}), we see the Einstein
equation is not consistent.

To solve this problem we introduce a Lorentz violating term in the gravity sector simultaneously. There are many
Lorentz violating modifications of gravity discussed in the literature. In this paper, we only consider the Chern-Simons modification proposed by
Jackiw and Pi \cite{Jackiw:2003pm}. So the total Lagrangian is
\be\label{lagrangian}
\mathcal{L}=\frac{1}{16\pi G}(R+\frac{\theta_2}{4} R\widetilde{R})-\frac{1}{4}F_{\mu\nu}F^{\mu\nu}
+\frac{\theta_1}{4}F_{\mu\nu}\widetilde{F}^{\mu\nu}+\mathcal{L}_m~,
\ee
where $R\widetilde{R}\equiv \frac{1}{2}\epsilon^{\mu\nu\alpha\beta}R^{\sigma}_{~\rho\alpha\beta}R^{\rho}_{~\sigma\mu\nu}$ is the
gravitational Chern-Pontryagin density and $R^{\sigma}_{~\rho\alpha\beta}$ is the Riemann tensor. Similarly the parameter $\theta_2=q_{\mu}x^{\mu}$,
where $q_{\mu}$ is a constant vector with dimension $[E^{-1}]$.
The variation of the action $S=\int d^4x \sqrt{-g}\mathcal{L}$ with respect to the metric gives the modified
Einstein equation \cite{Jackiw:2003pm,Alexander:2009tp}
\be\label{equation}
G^{\mu\nu}+C^{\mu\nu}=-8\pi G(T^{\mu\nu}_F+T^{\mu\nu}_m)~,
\ee
where
\be\label{cotten}
C^{\mu\nu}=-\frac{1}{2}[\nabla_{\sigma}\theta_2(\epsilon^{\sigma\mu\alpha\beta}\nabla_{\alpha}R^{\nu}_{~\beta}+
\epsilon^{\sigma\nu\alpha\beta}\nabla_{\alpha}R^{\mu}_{~\beta})+
\frac{1}{2}\nabla_{\sigma}\nabla_{\tau}\theta_2(\epsilon^{\sigma\nu\alpha\beta}R^{\tau\mu}_{~~\alpha\beta}+
\epsilon^{\sigma\mu\alpha\beta}R^{\tau\nu}_{~~\alpha\beta})]
\ee
is the four dimensional Cotten tensor, and its divergence gives
\be
\nabla_{\mu}C^{\mu\nu}=\frac{1}{8}q^{\nu} R\widetilde{R}~.
\ee
So, the divergence of Eq. (\ref{equation}) gives the constraint
\be\label{constraint}
q^{\nu}R\widetilde{R}=16\pi G p^{\nu} F_{\mu\alpha}\widetilde{F}^{\mu\alpha}~.
\ee

The normal Einstein field equation is a second order partial differential equation. When modified by the
gravitational Chern-Simons term, the field equation is promoted to third order
due to the four dimensional Cotten tensor shown in equation (\ref{cotten}). This means the constraint (\ref{constraint})
will not make the whole system overdetermined. But the solutions will be different from those obtained in general relativity. For
example, in Ref. \cite{Jackiw:2003pm}, the authors pointed out in the case $F_{\mu\alpha}\widetilde{F}^{\mu\alpha}=0$ that the Schwarzschild
solution is still existent but the Kerr solution is not.

\section{The effects on CMB anisotropies}

In the previous section, we considered in Eq. (\ref{lagrangian}) the Chern-Simons modified electromagnetic field with fixed preferred frame and
introduced the gravitational Chern-Simons term to modify the gravity sector simultaneously to preserve the consistency of the theory.
In this section we will
study the phenomenology of this model on CMB. For keeping the rotation invariance, we assume only the temporal components of $q_{\mu}$
and $p_{\mu}$ are not vanished.
During inflation the density $F_{\mu\nu}\widetilde{F}^{\mu\nu}$ is diluted and can be set to zero. So the gravitational Chern-Simons term
is constrained to be $R\widetilde{R}=0$. But at the linear order, as shown in Refs. \cite{Lue:1998mq,Jackiw:2003pm}, the gravity is modified
and the produced tensor perturbations have different intensities for different helicities.
The gravitational wave has two independent polarized components denoted by $+$ and $\times$. Here it is more
convenient to use the right- and left-handed circular polarized components:
\bea
& & h^{R}=\frac{1}{\sqrt{2}}(h^{+}-ih^{\times})~,\nonumber\\
& & h^{L}=\frac{1}{\sqrt{2}}(h^{+}+ih^{\times})~. \eea The
primordial gravitational waves generated during inflation freeze
out after exiting the horizon. The power spectra for different
handedness are defined as follows: \be \langle h^{R \ast}({\bf
k_1})h^R({\bf k_2})\rangle=P^R_h\delta^3({\bf k_1}-{\bf k_2})~,~
\langle h^{L\ast}({\bf k_1})h^L({\bf
k_2})\rangle=P^L_h\delta^3({\bf k_1}-{\bf k_2})~,~ \langle
h^{R\ast}({\bf k_1})h^L({\bf k_2})\rangle=0~. \ee As mentioned
above, due to the gravitational Chern-Simons term, the produced
$P^R_h$ and $P^L_h$ are not equal. The discrepancy depends on $q_0$ and the Hubble constant $H_{\rm in}$ during inflation and can be denoted by
the small parameter $\epsilon=-(\pi/2)q_0 H_{\rm in}$ \cite{Alexander:2004wk}: \be\label{PTprim}
P^R_h=\frac{1}{2}P_h(1-\epsilon)~,~P^L_h=\frac{1}{2}P_h(1+\epsilon)~,~P^R_h+P^L_h=P_h~,~P^R_h-P^L_h=-\epsilon
P_h~. \ee

As is well known, the primordial scalar perturbations can generate
the temperature and $E$-mode polarization perturbations in CMB.
The tensor perturbations can generate $B$-mode perturbations
besides $T$ and $E$. In the absence of the Chern-Simons term
the generated $B$-modes are not correlated with $T$ and $E$. As
mentioned above, with the gravitational Chern-Simons term, the non-vanished $TB$ and $EB$ correlations would be produced
at the last scattering surface, which we will explain
in detail in the following.

The polarization of electromagnetic field is described by the Stokes parameters $I$, $Q$, $U$ and $V$. For CMB physics, the Stokes $V$ is usually
neglected because the Thomson scattering cannot produce net circular polarizations. The intensity $I$ is invariant under coordinate transformations,
but $Q$ and $U$ are not. The combinations $Q\pm iU$ behave like spin-2 variables under the rotation. Given a map of temperature and polarization, we can
expand the perturbations in terms of spin-weighted harmonic function as below
\bea
T(\hat{\bi{n}})&=& \sum_{lm}a_{T,lm}Y_{lm}(\hat{\bi{n}})\nonumber \\
(Q\pm iU) (\hat{\bi{n}})&=& \sum_{lm} a_{\pm 2, lm} \;_{\pm 2}Y_{lm}(\hat{\bi{n}})~.
\eea
The expressions for the expansion coefficients are
\begin{eqnarray}
a_{T,lm}&=&\int d\Omega\; Y_{lm}^{*}(\hat{\bi{n}}) T(\hat{\bi{n}})
\nonumber  \\
a_{\pm 2,lm}&=&\int d\Omega \;_{\pm 2}Y_{lm}^{*}(\hat{\bi{n}}) (Q\pm iU)(\hat{\bi{n}})~.\label{alm}
\end{eqnarray}
Instead of $a_{2,lm}$ and $a_{-2,lm}$, it is convenient to introduce their
linear combinations
\begin{eqnarray}
a_{E,lm}=-(a_{2,lm}+a_{-2,lm})/2 \nonumber \\
a_{B,lm}=i(a_{2,lm}-a_{-2,lm})/2.
\label{aeb}
\end{eqnarray}
The power spectra are defined as \be \langle a_{X',l^\prime
m^\prime}^{*} a_{X,lm}\rangle= C^{X'X}_{l} \delta_{l^\prime l}
\delta_{m^\prime m} \ee with the assumption of statistical isotropy, where $X$ and $X'$ stand for $T$, $E$ and $B$.

In real space, using the spin raising and lowering operators $\edth$ and $\baredth$,  it is useful to introduce
two scalar quantities $\tilde{E}(\hat{\bi{n}})$ and $\tilde{B}(\hat{\bi{n}})$
defined as \cite{Zaldarriaga:1996xe}
\begin{eqnarray}
\tilde{E}(\hat{{\bi n}})&\equiv&
-{1\over 2}\left[\baredth^2(Q+iU)+\edth^2(Q-iU)\right]
\nonumber \\
    &=&\sum_{lm}\left[{(l+2)! \over (l-2)!}\right]^{1/2}
a_{E,lm}Y_{lm}(\hat{{\bi n}}) \nonumber \\
\tilde{B}(\hat{\bi n})&\equiv&{i\over 2}
\left[\baredth^2(Q+iU)-\edth^2(Q-iU)\right]
\nonumber \\
    &=&\sum_{lm}\left[{(l+2)! \over (l-2)!}\right]^{1/2}
a_{B,lm}Y_{lm}(\hat{\bi n})~.
\label{EBexpansions}
\end{eqnarray}

For each Fourier component, we can simply work in the coordinate frame in which $\hat{\bi{k}} \parallel \hat{\bi{z}}$
and then integrate over all the Fourier modes.
The generated temperature and polarization perturbations by gravitational waves can be expressed as \cite{Zaldarriaga:1996xe}:
\begin{eqnarray}
\Delta_T^{(T)}(\tau_0,\hat{\bi n},{\bi k}) &=&
\left[(1-\mu^2)e^{2i\phi}h^R({\bi k})
+ (1-\mu^2)e^{-2i\phi}h^L({\bi k})\right]
\tilde{\Delta}^{(T)}_T(\tau_0,\mu,k) \nonumber \\
(\Delta_Q^{(T)}\pm i\Delta_U^{(T)})
(\tau_0,\hat{\bi n},{\bi k}) &=& \left[(1\mp\mu)^2 e^{2i\phi}h^R({\bi k})
+ (1\pm\mu)^2e^{-2i\phi}h^L({\bi k})\right]
\tilde{\Delta}^{(T)}_P(\tau_0,\mu,k)~,
\label{integsolten}
\end{eqnarray}
where the superscript $T$ denotes tensor perturbations and $\tilde{\Delta}$ are the obtained Polnarev variables \cite{Polnarev}
by integrating the Boltzmann equations. We used the conformal time $\tau$ and $\tau_0$ to denote present value,
$x=k(\tau_0-\tau)$ and $\mu=\hat{\bi k}\cdot \hat{\bi n}$.
The quantities $\Delta_T^{(T)}$,
$\Delta_{\tilde{E}}^{(T)}$ and
$\Delta_{\tilde{B}}^{(T)}$ are given by
\begin{eqnarray}
\Delta_T^{(T)}
(\tau_0,\hat{\bi{n}},\bi{k})&=&\Big[(1-\mu^2)e^{2i\phi}h^R(\bi{k})+
(1-\mu^2)e^{-2i\phi}h^L(\bi{k})\Big]
\tilde{\Delta}^{(T)}_T(\tau_0,\mu,k)
\nonumber \\
\Delta_{\tilde{E}}^{(T)}
(\tau_0,\hat{\bi{n}},\bi{k})&=&\Big[(1-\mu^2)e^{2i\phi}h^R(\bi{k})+
(1-\mu^2)e^{-2i\phi}h^L(\bi{k})\Big]{\hat{\cal E}}(x)
\tilde{\Delta}^{(T)}_P(\tau_0,\mu,k)
\nonumber \\
\Delta_{\tilde{B}}^{(T)}
(\tau_0,\hat{\bi{n}},\bi{k})&=&\Big[(1-\mu^2)e^{2i\phi}h^R(\bi{k})-
(1-\mu^2)e^{-2i\phi}h^L(\bi{k})\Big]{\hat{\cal B}}(x)
\tilde{\Delta}^{(T)}_P(\tau_0,\mu,k)~,
\label{tebT}
\end{eqnarray}
where the operators are defined as
${\hat{\cal E}}(x)=-12+x^2[1-\partial_x^2]-8x\partial_x $ and
${\hat{\cal B}}(x)=8x+2x^2\partial_x$.
  From these equations
one can show that the spectra $TT$, $EE$, $BB$ and $TE$ only depend on the sum of
the primordial spectra of gravitational waves, $P_h=P^R_h+P^L_h$. But the cross correlations $TB$ and $EB$
relies on the difference $P^R_h-P^L_h=-\epsilon P_h$,
\begin{eqnarray}
C^{XX(T)}_l&=&
(4\pi)^2\int k^2dkP_h(k)\Big[\Delta^{(T)}_{Xl}(k)\Big]^2
\nonumber \\
C^{TE(T)}_l&=&
(4\pi)^2\int k^2dkP_h(k)\Delta^{(T)}_{Tl}(k)
\Delta^{(T)}_{El}(k)\nonumber\\
C^{TB(T)}_l&=&-
(4\pi)^2\epsilon\int k^2dkP_h(k)\Delta^{(T)}_{Tl}(k)
\Delta^{(T)}_{Bl}(k)\nonumber\\
C^{EB(T)}_l&=&-
(4\pi)^2\epsilon\int k^2dkP_h(k)\Delta^{(T)}_{El}(k)
\Delta^{(T)}_{Bl}(k)~.
\label{est}
\end{eqnarray}
The transfer functions $\Delta^{(T)}_{Xl}(k)$ are defined in Eq. (30) of Ref. \cite{Zaldarriaga:1996xe}.
We can see that the effect of gravitational Chern-Simons term on CMB is only to generate the $TB$ and $EB$ correlations and leave other
power spectra unmodified. This result is consistent with that of Ref. \cite{Saito:2007kt}.

The scalar perturbations also have contributions to $TT$, $EE$ and $TE$. So, the total power spectra of CMB should be
\bea
& & C^{TT}_l=C^{TT(S)}_l+C^{TT(T)}_l~,~C^{EE}_l=C^{EE(S)}_l+C^{EE(T)}_l~,~C^{TE}_l=C^{TE(S)}_l+C^{TE(T)}_l~,\nonumber\\
& & C^{BB}_l=C^{BB(T)}_l~,~C^{TB}_l=C^{TB(T)}_l~,~C^{EB}_l=C^{EB(T)}_l~.
\eea
After the perturbations of CMB generated at the last scattering surface, the polarizations of CMB photons are rotated under the influence of
the electromagnetic Chern-Simons term. The rotation angle is $\Delta\chi=(p_0/2)(\tau_0-\tau_{ls})$ \cite{Li:2008tma}, where $\tau_{ls}$
is the conformal time of the last scattering.  In this case, except $TT$, all other spectra are changed \cite{Feng:2006dp}:
\bea\label{rotationformulas}
 C^{TT,obs}_l&=&C^{TT}_l\nonumber\\
  C^{TE,obs}_l&=&C^{TE}_l \cos{(2\Delta\chi)}-C^{TB}_l \sin{(2\Delta\chi)} \nonumber\\
 C^{TB,obs}_l&=&C^{TE}_l \sin{(2\Delta\chi)}+C^{TB}_l \cos{(2\Delta\chi)}\nonumber\\
 C^{EE,obs}_l&=&C^{EE}_l \cos^2{(2\Delta\chi)} +C^{BB}_l \sin^2{(2\Delta\chi)}-C^{EB}_l\sin{(4\Delta\chi)} \nonumber\\
   C^{BB,obs}_l&=&C^{EE}_l \sin^2{(2\Delta\chi)} +C^{BB}_l \cos^2{(2\Delta\chi)}+C^{EB}_l\sin{(4\Delta\chi)} \nonumber\\
  C^{EB,obs}_l&=&\frac{1}{2}\sin{(4\Delta\chi)} (C^{EE}_l-C^{BB}_l)+C^{EB}_l\cos{(4\Delta\chi)} ~.
  \eea
In the formulas above, the quantities with the superscript $obs$
are those observed after the rotation. We can see that, in the standard case without the Lorentz and $CPT$ violations, only $TT$, $TE$, $EE$ and
$BB$ in the right hand side of Eqs. (\ref{rotationformulas})are produced. Consider the modifications, the gravitational Chern-Simons produce non-vanished
$TB$ and $EB$ in the right hand side of Eqs. (\ref{rotationformulas}), which are calculated by Eqs. (\ref{est}). Then the electromagnetic Chern-Simons
term further rotated the $E$ mode to $B$ mode polarization and change all but $TT$ spectra as showed in the left hand side of Eqs. (\ref{rotationformulas}).

\section{Energy spectrum of GWB at late-time evolution}

In this section we focus on the dynamics of gravitational waves at
late-time evolution. As shown in above section, the Chern-Simons
term does not contribute on the total primordial power spectrum of
tensor fluctuations but only affect their propagations.
Correspondingly, we expect the energy spectrum of GWB would be
modified by the Chern-Simons term.

We start by giving the equation of motion for the tensor
fluctuations in Fourier space\cite{Alexander:2004wk},
\begin{eqnarray}\label{eqomh}
(1-\lambda^sq_0\frac{k}{a}) {h^s_k}'' + 2{\cal
H}(1-\frac{1}{2}\lambda^sq_0\frac{k}{a}){h^s_k}'
+(1-\lambda^sq_0\frac{k}{a})k^2{h^s_k}=16\pi Ga^2\sigma^s_k~,
\end{eqnarray}
where the prime denotes the derivative with respect to the
conformal time $\tau\equiv\int^\tau\frac{dt}{a(t)}$. The subscript
$s$ represents the two polarizations, with $\lambda^R=1$ and
$\lambda^L=-1$.  $\sigma^{s}$ is the anisotropic part of the
stress tensor, constructed by the spatial components of the
perturbed energy-momentum tensor. We would like to neglect it
first, and then consider its contribution on transfer function
later.

\subsection{Canonical representation}

To simplify the equation of motion for tensor fluctuations, we
introduce a quantity $\nu^s_k$\cite{Choi:1999zy}, which is given
by
\begin{eqnarray}
\nu^s_k(\tau) \equiv a\sqrt{1-\lambda^sq_0\frac{k}{a}}~.
\end{eqnarray}
Then we redefine the variables of fluctuations as
\begin{eqnarray}
v_k^s\equiv \nu^s_kh^s_k~,
\end{eqnarray}
which are often viewed as generalized Mukhanov-Sasaki variables.
The equation of motion for $v^s_k$ is
given by
\begin{eqnarray}\label{eom}
{v^s_k}''+(k^2-\frac{{\nu^s_k}''}{\nu^s_k})v^s_k=0~.
\end{eqnarray}

Eq. (\ref{eom}) has an asymptotic solution when we neglect the
last term $\frac{{\nu^s_k}''}{\nu^s_k}$ which implies
$|k\tau|\gg1$, and it is strongly oscillating like trigonometric
functions. This feature coincides with an adiabatic condition,
which corresponds to the case that the effective physical
wavelength is deep inside the Hubble radius. Therefore, the modes
can be regarded as adiabatic when they are staying in the
sub-Hubble regime with $|k\tau|\gg1$, and we may impose a suitable
initial condition in virtue of WKB approximation,
\begin{eqnarray}\label{inicond}
v^s_k\simeq\frac{1}{\sqrt{2k}}e^{-ik\tau}~,
\end{eqnarray}
for cosmological fluctuations.

Once we have resolved the solutions to the above equations, we can
obtain the tensor power spectrum $P_h^s$ for the polarization mode
$h^s_k$. The GWB we may observe today should be characterized by
the energy spectrum, defined by
\begin{eqnarray}\label{tensor energy}
\Omega_{GW}(k,
\tau)\equiv\frac{1}{\rho_c(\tau)}\frac{d\langle0|\rho_{GW}(\tau)|0\rangle}{d\,{\rm
ln}\,k}~,
\end{eqnarray}
where $\rho_{GW}(\tau)$ indicates the energy density of
gravitational waves, and the parameter $\rho_c(\tau)$ is the
critical density of the universe. Since the GWB we
observed has already re-entered the horizon, its mode should
oscillate in the form of sinusoidal function. Accordingly, we can
deduce the relation between the power spectrum and the energy
spectrum at the scales of interests as follows
\begin{eqnarray}\label{approximate tensor energy}
\Omega_{GW}(k,\tau)\simeq\frac{1}{12}\frac{k^2}{a^2(\tau)H^2(\tau)}\sum_s
P_h^s(k, \tau)~,
\end{eqnarray}
where we have used the Friedmann equation $H^2(\tau)=\frac{8\pi
G}{3}\rho_c(\tau)$.

\subsection{Transfer function}

Now we analyze the evolution of tensor perturbations in the GWB
nowadays. Since the primordial gravitational waves are distributed
in every frequency, once the effective co-moving wave number is
less than $aH$, the corresponding mode of gravitational waves
would escape the horizon and be frozen until it re-enters the
horizon. The relation between the time when tensor perturbations
leave the horizon and the time when they return is
$a_{out}H_{out}=a_{in}H_{in}$. Therefore, we have the conclusion
that, the earlier the perturbations escape the horizon, the later
they re-enter it. Moreover, once the effective co-moving wave
number is larger than $aH$, the perturbations begin to oscillate
like the plane wave. In the following, we will establish the
relation to relate the power spectrum observed today to the
primordial one. It can be described by the transfer functions
which are defined as follows,
\begin{eqnarray}\label{PTtoday}
P_h^s(k,\tau)=T^s(k,\tau)P_h^s(k,\tau_i)~,
%\\\label{PTtoday2}P_h^R(k,\tau)=T^R(k,\tau)P_h^R(k,\tau_i)~,
\end{eqnarray}
where $\tau_i$ indicates the end of primordial inflation.

In order to make clear every possible ingredient affecting the
evolvement of the GWB, it is suitable and reasonable to decompose
the transfer function into three parts as
follows\cite{Boyle:2005se},
\begin{eqnarray}
T^s(k,\tau)&=&F^s_1F^s_2F^s_3\nonumber\\
&=&|\frac{\bar h^s_k(\tau)}{h^s_k(\tau_i)}|^2|\frac{\tilde
h^s_k(\tau)}{\bar h^s_k(\tau)}|^2|\frac{h^s_k(\tau)}{\tilde
h^s_k(\tau)}|^2~.
\end{eqnarray}
In the above formula, $h^s_k(\tau)$ is the exact solution of Eq.
(\ref{eqomh}); $\tilde h^s_k(\tau)$ is an approximate solution of
Eq. (\ref{eqomh}) by neglecting the anisotropic stress tensor; and $\bar h^s_k(\tau)$ is also an approximate
solution which is equal to $h^s_k(\tau_i)$ if $k<aH$ while equal
to plane wave if $k>aH$.

First, from Eq. (\ref{inicond}) one can see that after horizon
re-entering, gravitational waves begin to oscillate with a
decaying amplitude proportional to $1/\nu^s_k(\tau)$. Therefore,
from the definition of $\bar h^s_k$ we get
\begin{eqnarray}
  \bar h^s_k(\tau)= \left\{ \begin{array}{c}
    \frac{A^s_k}{\nu^s_k(\tau)}\cos[k(\tau-\tau_k)+\phi^s],~~k>aH \\
    \\
    h^s_k(\tau_i),~~k<aH
\end{array} \right.  \label{baru1}
\end{eqnarray}
where $\phi_l$ depends on the initial condition, $A^s_k$ is the
maximum of the amplitude of oscillations, and $\tau_k$ is the
conformal time when $k=aH$. Since we require this function to be
continuous, there must be a matching relation that
$h^s_k(\tau_i)=[A^s_k\cos\phi^s]/\nu^s_k(\tau_k)$. Based on these
relations one can get the first factor $F^s_1$ as follows
\begin{eqnarray}\label{F1L}
F^s_1=\left(\frac{\nu^s_k(z_k)}{\nu^s_k(z)}\right)^2\cos^2[k(\tau-\tau_k)+\phi^s]/\cos^2\phi^s,
\end{eqnarray}
where we introduce the redshift $1+z=a_0/a(\tau)$ in aim of
showing the effects of the suppression as a result of redshift.
The index ``$0$" indicates today, and $z_k$ is the redshift when
the modes re-entered the horizon $k=aH$. Note that the relation of
$z_k$ and $k$ can be given by the following equation
\begin{eqnarray}
\left(\frac{k}{k_0}\right)^2=\sum_i\Omega_i^{(0)}(1+z_k)\exp\left[3\int_0^{z_k}\frac{w_i(\tilde
z)}{1+\tilde z}d\tilde z \right],
\end{eqnarray}
where the sum over $i$ includes all components in the universe.
Since the contributions from dark energy and the fluctuations in
radiation are very small, here we ignore them and then get
\begin{eqnarray}\label{zl}
1+z_k=\frac{1+z_{eq}}{2}\left[-1+\sqrt{1+\frac{4(k/k_0)^2}{(1+z_{eq})\Omega_m^{(0)}}}\right],
\end{eqnarray}
where $z_{eq}\equiv-1+\Omega_m^{(0)}/\Omega_r^{(0)}$. The factor
$F^s_1$ describes the redshift-suppressing effect on the
primordial gravitational waves. Since this factor shows strongly
oscillating behaviour which is inconspicuous to be observed in the
GWB, we usually average the term $\cos^2[k(\tau-\tau_k)+\phi^s]$
and obtain $\frac{1}{2}$ for it.

Second, when considering the influence of the background equation of state of
universe on the re-entry of horizon, we focus on analyzing the
factor $F^s_2$. Since the background equation of state $w$ varies
very slowly, it is profitable to assume that the evolution of the
scale factor is of form $a=a_0(\frac{\tau}{\tau_0})^{\alpha}$ with
$\alpha=\frac{2}{1+3w}$. To ignore the anisotropic stress tensor
$\sigma^s$ and the Chern-Simons modifications, we resolve Eq.
(\ref{eqomh}) again and then have
\begin{eqnarray}
\tilde
h^s_k(\tau)=h^s_k(\tau_i)\Gamma(\alpha+\frac{1}{2})\left(-\frac{k\tau}{2}\right)^{\frac{1}{2}-\alpha}J_{\alpha-\frac{1}{2}}(-k\tau)~,
\end{eqnarray}
where $\Gamma$ is the Gamma function and $J_{\nu}$ is the $\nu$-th
Bessel function. If $|k\tau|\gg1$, there is such a relation that
$|\frac{\tilde
u_1(k,\tau)}{u_1(k,\tau_i)}|^2=\frac{\Gamma^2(\alpha+\frac{1}{2})}{\pi}(-\frac{k\tau}{2})^{-2\alpha}\cos^2(k\tau+\frac{\alpha\pi}{2})$.
To match with Eq. (\ref{F1L}), considering that the phase should
be continuous, hence we have the solution that when tensor
fluctuations re-enter the horizon the conformal time
$\tau_k=-\frac{\alpha}{k}$. A similar relation was obtained in
\cite{Cai:2008ed} during primordial stage. Finally,
the second factor $F^s_2$ is given by
\begin{eqnarray}\label{F2L}
F^s_2=\frac{\Gamma^2(\alpha+\frac{1}{2})}{\pi}\left(\frac{2}{\alpha}\right)^{2\alpha}\cos^2\phi^s~.
\end{eqnarray}
The second factor shows that, when the gravitational waves
re-enter the horizon, there is a "wall" lying on the horizon which
affects the tensor power spectrum.

Third, during the evolution of tensor perturbations, the nonzero
anisotropic stress tensor $\sigma^s$ would more or less bring some
effects on the GWB. This effect is pointed out by Steven
Weinberg\cite{Weinberg:2003ur}, and usually the primary ingredients
are the freely streaming neutrinos which damp the amplitude of the
tensor power spectrum. This damping effect just makes power spectrum
of tensor fluctuation times a constant but do not change the
dynamics of the GWB's evolution. A combination of analytic and
numerical calculations performed in Refs. \cite{Pritchard:2004qp,
Dicus:2005rh, Zhao:2009we,Zhao:2006mm,Giovannini:2009kg} suggests
that $F^s_3=0.80313$ for the frequency of relic gravitational waves
among $10^{-16}Hz$ and $10^{-10}Hz$ is in high precision.

Eventually, we have discussed three kinds of leading corrections
in the transfer function which make contributions in the evolution
of the GWB. One can see that the modifications brought by the
Chern-Simons term contribute mostly to the first factor $F^s_1$. This is because the Chern-Simons
term of gravity sector mainly affect the physics at high energy
scale.

\subsection{Analysis of today's GWB}

Using the transfer functions, we are able to connect the
primordial gravitational waves with what we observe today. To
substitute Eqs. (\ref{F1L}), (\ref{F2L}), and the damping factor
$F_3$ into (\ref{PTtoday}), and making use of the primordial
tensor spectrum (\ref{PTprim}) we can give today's tensor power
spectrum as follows,
\begin{eqnarray}\label{tensor power today}
P^s_h(k,\tau_0)&=&T^s(k,\tau_0)P^s_h(k,\tau_i)\nonumber\\
&=& \frac{1-\lambda^sq_0k(1+z_k)}{1-\lambda^sq_0k}
\frac{F_3^2}{(1+z_k)^2} \frac{\Gamma^2(\alpha+\frac{1}{2})}{2\pi}
\bigg(\frac{2}{\alpha}\bigg)^{2\alpha} \times P_h^s~.
\end{eqnarray}

\begin{figure}
\begin{center}
\includegraphics[width=3.9in]{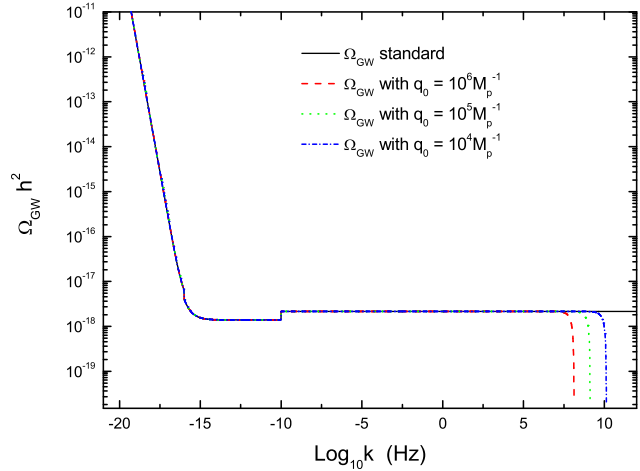}
\end{center}
\caption{The plot of energy spectra of relic gravitational waves.
The black solid line represents the energy spectrum without
Chern-Simons modification; the red dash line gives the curve of
energy spectrum with $q_0=10^{6}M_p^{-1}$; the green dot curve
shows the energy spectrum with $q_0=10^{5}M_p^{-1}$; and the blue
dot-dash line shows the energy spectrum with $q_0=10^{4}M_p^{-1}$.
Here the potential of inflation is taken to be $V_{inf}\simeq M^4$
in which $M=5\times10^{15}$GeV and $M_p$ is the reduced Planck
mass.} \label{fig:Omega}
\end{figure}

When the frequency of GWB is small enough, the time for the
corresponding mode re-entering the horizon is close to today, and
one can see that the tensor perturbation power spectrum would
agree with the standard theory very well. Therefore, we expect to
find signals in the high-frequency region. To make a comparison
with the normal energy spectrum of relic gravitation waves, we
assume the slow-roll parameter approaches zero during
inflation and the potential of the inflaton is $V_{inf}\sim M^4$
with the scale $M=5\times 10^{15}GeV$. Consequently we
can obtain the semi-analytical form of the present energy spectrum
of tensor perturbations with Chern-Simons modifications as
follows,
\begin{eqnarray}
\Omega_{GW}(k,\tau_0)h^2 = \frac{gk^2}{(1-\sqrt{1+fk^2})^2} \times
\sum_s(1-\lambda^s\epsilon)\frac{1-\lambda^sq_0k(1+z_k)}{1-\lambda^sq_0k}~,
\end{eqnarray}
where the numerical calculations show $f=3.10475\times10^{32}$,
and $g$ equals to $2.15691\times10^{14}$ when the frequency is
between $10^{-16}$ and $10^{-10}Hz$ but takes the value
$3.34395\times10^{14}$ outside this region. In the numerical
computation, we have taken $w=-1$ and $a_0=1$. We provide the
numerical results in Fig. \ref{fig:Omega}.

From Fig. \ref{fig:Omega}, one can see that the energy spectra of
relic gravitational waves with Chern-Simons term coincide with
that in standard gravity theory at low-frequency regime, but are
suppressed at high-frequency regime. Moreover, the suppressions of
energy spectra strongly depend on the value of the parameter
$q_0$. A similar suppression effect on energy spectrum of GWB due
to a spacetime non-commutativity was found in \cite{Cai:2007xr},
but it takes place at low-frequency regime. We can understand this
effect as follows. A Lorentz-violating term often brings an
effective mass for gravitons which could suppress the energy
spectrum. Therefore, this suppression depends on the energy scale
of the effective mass term. Although for both
non-commutativity and Chern-Simons modifications, the Lorentz
violations take place at high-energy scales, the effective mass
caused by non-commutativity appears at low-frequency regime but
that brought by Chern-Simons term happens at high-frequency
regime.

\section{Summary and Conclusions}

In this paper, we pointed out that the Lorentz violations induced by fixed preferred frames in the matter sector
will make the gravitational field equation inconsistent because the energy-momentum tensor cannot be both symmetric and covariantly
conserved. We provided a solution to this problem by modifying the gravity simultaneously.
In a concrete model, we considered the Chern-Simons modified electrodynamics which breaks Lorentz and $CPT$ invariance.
Simultaneously the gravity is also modified by a Chern-Simons term. The phenomenologies of this model on CMB and the late-time dynamics of
relic gravitational waves have been studied. For these modifications, the gravitational Chern-Simons term generates
non-vanished $TB$ and $EB$ cross-correlations of CMB at the last scattering surface and leave others unchanged. After that, the
electromagnetic Chern-Simons term will rotate the generated $E$-mode polarizations to $B$-mode ones and change all but $TT$ spectra.
For the late-time evolution of the relic gravitational waves, we found that the Chern-Simons term mainly
contribute to the amplitude of GWB inside the horizon. In this case, the energy spectrum of GWB is suppressed at
high-frequency regime, which depends on an effective mass brought by the Chern-Simons term. From the current result, we
notice that this effect is hard to be detected by recent experiments. However, the studies on gravitational models with
various Lorentz violations \cite{Alexander:2004us, Cai:2009in} and their gravitational perturbations \cite{Cai:2009hc, Satoh:2007gn}
are particularly of theoretical interests.
This model deserves further studies. It will be interesting to seek for
Kerr-Newmann type solutions for the spacetime metric in the region surrounding a charged and rotating celestial body.

In the literature, there is another class of Lorentz violating theories in which the Lorentz symmetry is broken spontaneously,
for example, the external vector is replaced by the derivative of a dynamical scalar field, $p_{\mu}=\nabla_{\mu}\phi$.
In this case, we should include the kinetic term and the potential of $\phi$ in the action. During the evolution of the universe,
the scalar field develops a non-zero $\nabla_{\mu}\phi$ and the Lorentz symmetry is broken in this background.
Such Lorentz violation has been used to generate the baryon number asymmetry observed in our universe
\cite{Cohen:1987vi,Li:2001st,Davoudiasl:2004gf,Li:2004hh}. The scalar field may be the dynamical dark
energy \cite{Li:2001st} or the curvature scalar \cite{Davoudiasl:2004gf,Li:2004hh}. In this case, there is no problem of consistency and it
is not necessary to modify the gravity theory \cite{Kostelecky:2003fs,Bailey:2006fd}.
For example, in the Maxwell-Chern-Simons theory considered in Eq. (\ref{chernsimons}), if
$p_{\mu}=\nabla_{\mu}\phi$ and the kinetic term and potential of $\phi$ are included, the total energy-momentum tensor
$T^{\mu\nu}_F+T^{\mu\nu}_{\phi}$ is symmetric and divergence free even though $T^{\mu\nu}_F$ and $T^{\mu\nu}_{\phi}$ are not covariantly
conserved individually. The divergence of $T^{\mu\nu}_F$ is canceled by that of $T^{\mu\nu}_{\phi}$.
An interesting effect in this case is that the rotation angle of the photon is dynamical and
spacetime dependent \cite{Li:2008tma}.

\begin{acknowledgments}
We would like to thank Jun-Qing Xia for discussions. The author
X.W. is supported by Starting Fund of Nanjing University under
Grant No. 0204003140. The researches of Y.F.C. and X.M.Z. are
supported in part by the National Science Foundation of China
under Grants No. 10533010 and 10675136, and by the Chinese Academy
of Sciences under Grant No. KJCX3-SYW-N2.
\end{acknowledgments}

\end{document}